\documentclass[12pt,reqno,a4paper]{amsart}
\usepackage{amsmath,amssymb,dsfont,graphicx,cite,latexsym,epsf,cancel}
\usepackage[colorlinks=false]{hyperref}
\newtheorem{theorem}{Theorem}

\newtheorem{definition}{Definition}

\makeatletter
\expandafter\let\expandafter
\reset@font\csname reset@font\endcsname
\def\subeqnarray{\arraycolsep1pt
   \def\@eqnnum\stepcounter##1{\stepcounter{subequation}
       {\reset@font\rm(\theequation\alph{subequation})}}
\jot5mm     \eqnarray}

\makeatother
\newcommand{\bbZ}{{\mathbb Z}}
\newcommand{\bbR}{{\mathbb R}}
\newcommand{\cL}{{\mathcal L}}
\def\epsilon{\varepsilon}

\def\tilde{\widetilde}

\def\endpf{$\blacksquare$\medskip}
\def\wx{\widetilde{x}}
\def\wip{\widetilde{p}}
\def\whx{\widehat{x}}
\def\whp{\widehat{p}}
\newbox\meibox
\def\placeunder#1#2#3#4{\setbox\meibox%
\vbox{\hbox{\hskip#4$\hphantom{#2}$}\hbox{$\hphantom{#1}$}}%
\vtop{\baselineskip=0pt\lineskiplimit=\baselineskip%
\lineskip=#3\hbox to \wd\meibox{\hfil\hskip#4$#2$\hfil}%
\hbox to \wd\meibox{\hfil$#1$\hfil}}}
\def\undertilde#1{\mathchoice{%
\placeunder{\vbox to 1.4pt{\hbox{$\displaystyle\widetilde{\,\,\,
}$}\vss}}{\displaystyle#1}{1.5pt}{1.5pt}}%
{\placeunder{\vbox to 1.4pt{\hbox{$\textstyle\widetilde{\,\,
}$}\vss}}{\textstyle#1}{1.5pt}{1.5pt}}%
{\placeunder{\vbox to 1.4pt{\hbox{$\scriptstyle\tilde{
}$}\vss}}{\scriptstyle#1}{1pt}{1pt}}%
{\placeunder{\vbox to 1.4pt{\hbox{$\scriptscriptstyle\tilde{
}$}\vss}}{\scriptscriptstyle#1}{1pt}{1pt}}%
}
\def\underhat#1{\mathchoice{%
\placeunder{\vbox to 1.4pt{\hbox{$\displaystyle\widehat{\,\,\,
}$}\vss}}{\displaystyle#1}{1.5pt}{1.5pt}}%
{\placeunder{\vbox to 1.4pt{\hbox{$\textstyle\widehat{\,\,
}$}\vss}}{\textstyle#1}{1.5pt}{1.5pt}}%
{\placeunder{\vbox to 1.4pt{\hbox{$\scriptstyle\widehat{
}$}\vss}}{\scriptstyle#1}{1pt}{1pt}}%
{\placeunder{\vbox to 1.4pt{\hbox{$\scriptscriptstyle\hat{
}$}\vss}}{\scriptscriptstyle#1}{1pt}{1pt}}%
}
\def\intprod{\mathbin{\hbox to 6pt{%
                 \vrule height0.4pt width5pt depth0pt
                 \kern-.4pt
                 \vrule height6pt width0.4pt depth0pt\hss}}}
\begin{document}
%%%%%%%%%%%%%%%%%%%%%%%%%%%%%%%
%%%%%%%%%%%%%%%%%%%%%%%%%%%%%%%
\title[Multi-time Lagrangian 1-forms]
{Variational formulation \\ of commuting Hamiltonian flows:\\ multi-time Lagrangian 1-forms}

\author{Yuri B. Suris }

\maketitle

\begin{center}
{\footnotesize{
Institut f\"ur Mathematik, MA 7-2,\\
Technische Universit\"at Berlin, Str. des 17. Juni 136\\
10623 Berlin, Germany\\
E-mail: {\tt suris@math.tu-berlin.de}
}}
\end{center}

%\tableofcontents

\begin{abstract} Recently, Lobb and Nijhoff initiated the study of variational (Lagrangian) structure of discrete integrable systems from the perspective of multi-dimensional consistency. In the present work, we follow this line of research and develop a Lagrangian theory of integrable one-dimensional systems. We give a complete solution of the following problem: one looks for a function of several variables (interpreted as multi-time) which delivers critical points to the action functionals obtained by integrating a Lagrangian 1-form along any smooth curve in the multi-time. The Lagrangian 1-form is supposed to depend on the first jet of the sought-after function. We derive the corresponding multi-time Euler-Lagrange equations and show that, under the multi-time Legendre transform, they are equivalent to a system of commuting Hamiltonian flows. Involutivity of the Hamilton functions turns out to be equivalent to closeness of the Lagrangian 1-form on solutions of the multi-time Euler-Lagrange equations. In the discrete time context, the analogous extremal property turns out to be characteristic for systems of commuting symplectic maps. For one-parameter families of commuting symplectic maps (B\"acklund transformations), we show that their spectrality property, introduced by Kuznetsov and Sklyanin, is equivalent to the property of the Lagrangian 1-form to be closed on solutions of the multi-time Euler-Lagrange equations, and propose a procedure of constructing Lax representations starting from the maps themselves.
\end{abstract}

%%%%%%%%%%%%%%%%%%%%%%%%%%%%%%%
%%%%%%%%%%%%%%%%%%%%%%%%%%%%%%%
\section{Introduction}
%%%%%%%%%%%%%%%%%%%%%%%%%%%%%%%
%%%%%%%%%%%%%%%%%%%%%%%%%%%%%%%

\begin{figure}[tp]
\begin{center}
\includegraphics[width=0.4\textwidth]{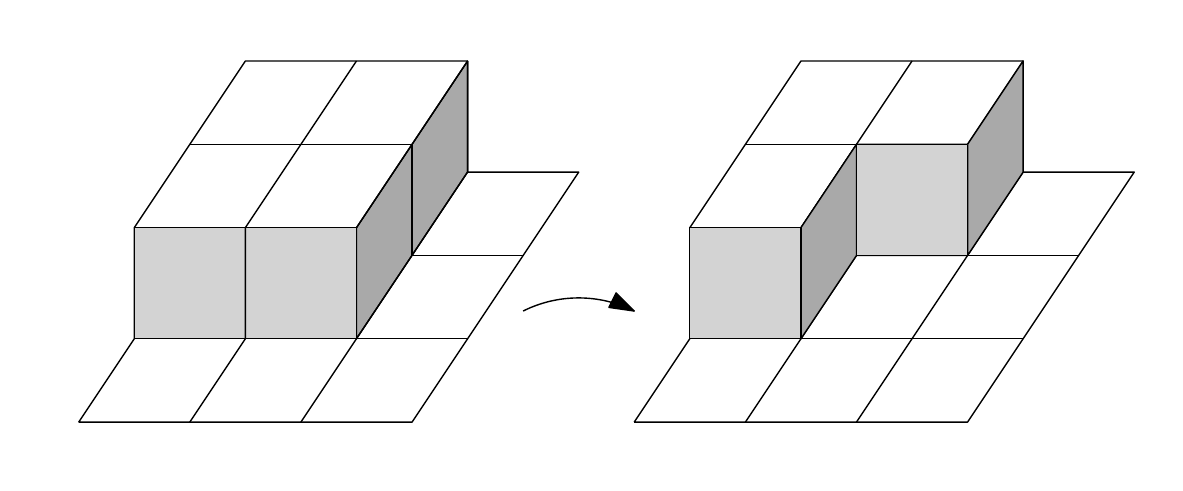}
\includegraphics[width=0.4\textwidth]{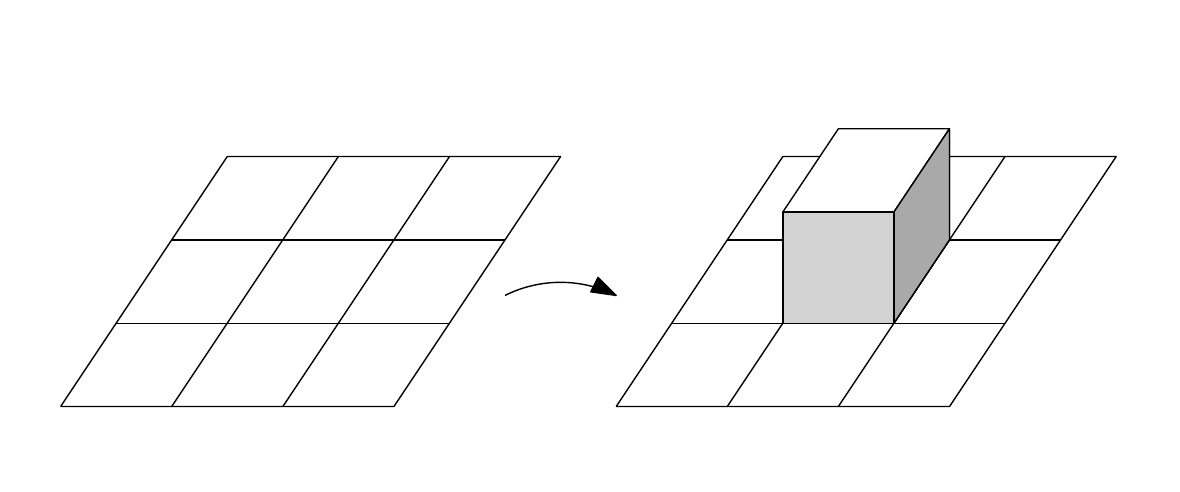}
\caption{Local moves of quad-surfaces involving one three-dimensional cube}
\label{Fig: local moves}
\end{center}
\end{figure}

It is by now well accepted that considering discrete systems brought a lot of new insight and a great deal of simplification into the general theory of integrable systems. This is most visible in the field of discrete differential geometry \cite{DDG}, but also on a more general side, a new understanding of integrability of discrete systems as their multi-dimensional consistency has been a major breakthrough \cite{BS1}, \cite{N}. This led to classification results \cite{ABS} about discrete 2-dimensional integrable systems (ABS list) which turned out to be rather influential. According to the concept of multi-dimensional consistency, integrable two-dimensional systems can be imposed in a consistent way on all two-dimensional sublattices of a lattices $\bbZ^m$ of arbitrary dimension. This means that the resulting multi-dimensional system possesses solutions whose restrictions to any two-dimensional sublattice are generic solutions of the corresponding two-dimensional system. To put this idea differently, one can impose the two-dimensional equations on any quad-surface in $\bbZ^m$ (i.e., a surface composed of elementary squares), and transfer solutions from one such surface to another one, if they are related by a sequence of local moves, each one involving one three-dimensional cube, like the moves shown of Fig. \ref{Fig: local moves}.

A further fundamental conceptual development was initiated by Lobb and Nijhoff \cite{LN1} and deals with variational (Lagrangian) formulation of discrete integrable systems. Solutions of any ABS equation on any quad surface are critical points of a certain action functional obtained by integration of a suitable discrete Lagrangian 2-form (i.e., a skew-symmetric real-valued function of an oriented elementary square). Lobb and Nijhoff observed that the value of the action functional remains invariant under local changes of the underlying quad-surface, and suggested to consider this as a defining feature of integrability. Their results, found on the case-by-case basis for some equations of the ABS list, have been extended to the whole list and given a more conceptual proof in \cite{BS2}, and have been subsequently generalized in various directions: for multi-field two-dimensional systems \cite{LN2}, for dKP, the fundamental three-dimensional discrete integrable system \cite{LNQ}, and, most recently, for the discrete time Calogero-Moser system, an important one-dimensional integrable system \cite{YLN}.

It is the latter generalization, concerned with the Lagrangian structure of integrable systems of classical mechanics, that we investigate in the present paper. While \cite{YLN} was mainly occupied with one concrete example, our aim here is to build of a general theory, both in the continuous and discrete time. In \cite{YLN}, an attempt of such a general theory for two-dimensional time was undertaken but unfortunately spoiled by a mistake\footnote{in their equation (8.11$b$), the derivatives $dt_2/ds$ and $dt_3/ds$ should be interchanged; as a consequence, their Euler-Lagrange equation (8.16) is incorrect.}.

In sections \ref{sect: cont results}, \ref{sect: cont proofs}, we develop the theory for any dimension of the time. The starting point is the following problem: under what conditions does a function $x$ of the multi-time $(t_1,\ldots,t_m)\in\bbR^m$ deliver critical points to functionals $$S_\Gamma=\int_\Gamma \sum_{\alpha=1}^m L_\alpha(x,x_{t_1},\ldots,x_{t_m})dt_\alpha$$ for any smooth curve $\Gamma$ in $\bbR^m$? At first glance, this seems to be a rather exotic problem. However (I am grateful to U. Pinkall and F. Pedit for the corresponding hint), a closely related problem lies behind an important theory of pluriharmonic functions and maps. Indeed, a pluriharmonic function of several complex variables minimizes the Dirichlet functional along any holomorphic curve in its domain. Not surprisingly, both our theory and the theory of pluriharmonic functions have one feature in common, namely their relation to integrable systems. We derive Euler-Lagrange equations for our problem and show that under the Legendre transform they turn into a system of commuting Hamiltonian flows with Hamilton functions $H_\alpha$. Moreover, the extremal property formulated above is characteristic for commuting Hamiltonian flows (i.e., for an arbitrary system of commuting Hamiltonian flows one can construct a Lagrangian 1-form with the above mentioned extremal property). Note that Hamiltonian flows commute if and only if the Poisson brackets of their Hamilton functions are constant, $\{H_\alpha,H_\beta\}=h_{\alpha\beta}={\rm const}$. Thus, this situation is not necessarily related to complete integrability in the Liouville-Arnold sense, where all Poisson brackets should vanish (see, e.g., \cite{RW} for a study of sets of Hamilton functions with constant Poisson brackets). It turns out that vanishing of the Poisson brackets $h_{\alpha\beta}=0$ is equivalent to the property of the Lagrangian 1-form to be closed on solutions of the Euler-Lagrange equations. In other words, for Liouville-Arnold integrable systems the value of action $S_\Gamma$ on the solutions is invariant under the local changes of the curve $\Gamma$.

In sections \ref{sect: discr results}, \ref{sect: discr proofs}, we develop the discrete counterpart of this theory. As usual, it turns out to be much more transparent and technically simple. Again, the extremal property for an arbitrary discrete curve in $\bbZ^m$ turns out to be characteristic for a system of $m$ commuting symplectic maps. For B\"acklund transformations, i.e., for one-parameter families of commuting symplectic maps, we show that existence of a certain generating function of common integrals of motion, given by the derivative of the Lagrange function with respect to the family parameter, is equivalent to closeness of the discrete Lagrangian 1-form. This gives a very nice interpretation and explanation of the spectrality property of B\"acklund transformations discovered by Kuznetsov and Sklyanin in \cite{KS}. On a more general note, it seems to be an important and difficult problem to find general enough conditions assuring integrability of a given system of commuting symplectic maps.

For B\"acklund transformations, we push forward the idea that such a standard integrability attribute as Lax or zero curvature representation (yielding, among other features, existence of a sufficient number of integrals of motion) can be extracted from the maps themselves. To put this more aphoristically, a family of B\"acklund transformations serves as its own Lax representation. This idea is conformal with the corresponding idea for discrete two-dimensional systems with the property of multi-dimensional consistency, which also serve as zero curvature representations for themselves (cf. \cite{BS1}, \cite{N}). In section \ref{sect: BT Toda}, we demonstrate, by the way of example, how our theory works for B\"acklund transformations for the classical Toda lattice.

%%%%%%%%%%%%%%%%%%%%%%%%%%%%%%%
%%%%%%%%%%%%%%%%%%%%%%%%%%%%%%%
\section{Basic construction in the continuous time case}
\label{sect: cont results}
%%%%%%%%%%%%%%%%%%%%%%%%%%%%%%%
%%%%%%%%%%%%%%%%%%%%%%%%%%%%%%%
Consider a 1-form on $\bbR^m$ (which we call the multi-dimensional time or simply multi-time), whose coefficients are functions on $X\times X^m$:
\begin{equation}
\cL(x,v_1,\ldots,v_m)=\sum_{\alpha=1}^m L_\alpha(x,v_1,\ldots,v_m) dt_\alpha.
\end{equation}
Here $X$ is the configuration space of a variational problem (usually we take $X=\bbR^N$).
We are looking for a function $x:\bbR^m\to X$ which solves the following variational problem. For any (smooth) curve $\Gamma:[0,1]\to\bbR^m$, it is required that $x$ delivers a critical point for the functional
\begin{equation}\label{eq: functional}
    S_\Gamma=\int_\Gamma\cL(x,x_{t_1},\ldots,x_{t_m}).
\end{equation}
We will prove the following results.
\begin{theorem}\label{th: EL}
The function  $x:\bbR^m\to X$ delivers a critical point for the functional $S_\Gamma$ for any smooth curve $\Gamma$, if and only if the following conditions (which we call {\em multi-time Euler-Lagrange equations}) are satisfied:
\begin{itemize}
\item For any $1\le \alpha\neq\beta\le m$,
\begin{equation}\label{eq: offdiag}
    \frac{\partial L_\alpha}{\partial v_\beta}(x,x_{t_1},\ldots,x_{t_m})=0;
\end{equation}
\item The following relations hold true:
\begin{equation}\label{eq: diag}
    \frac{\partial L_1}{\partial v_1}(x,x_{t_1},\ldots,x_{t_m})
    %=\frac{\partial L_2}{\partial v_2}(x,x_{t_1},\ldots,x_{t_m})
    =\ldots=\frac{\partial L_m}{\partial v_m}(x,x_{t_1},\ldots,x_{t_m}).
\end{equation}
We denote the common value of these functions by $p:\bbR^m\to X$;
\item This function $p$ satisfies the following differential equations:
\begin{equation}\label{eq: EL}
    \frac{\partial p}{\partial t_\alpha}=\frac{\partial L_\alpha}{\partial x}(x,x_{t_1},\ldots,x_{t_m}).
\end{equation}
\end{itemize}
\end{theorem}

The system of multi-time Euler-Lagrange equations is usually overdetermined, and it is not easy to analyze its compatibility in general. Indeed:
\begin{itemize}
\item Among $m(m-1)+(m-1)=m^2-1$ (vector) equations (\ref{eq: offdiag}), (\ref{eq: diag}), there should be precisely $m-1$ independent ones.
\item Differential equations (\ref{eq: EL}) are second order partial differential equations, and the same is true for their compatibility conditions,
\begin{equation}\label{eq: p compatible}
\frac{\partial}{\partial t_\alpha}\left(
\frac{\partial L_\beta}{\partial x}(x,x_{t_1},\ldots,x_{t_m})\right)-
\frac{\partial}{\partial t_\beta}\left(
\frac{\partial L_\alpha}{\partial x}(x,x_{t_1},\ldots,x_{t_m})\right)=0.
\end{equation}
One can expect that these compatibility conditions should be satisfied by virtue of equations (\ref{eq: EL}) themselves.
\end{itemize}
\medskip

{\bf Example.} Consider the two-time Lagrangian 1-form with the following components:
\begin{eqnarray}
L_1(x,x_{t_1}) & = & \frac{1}{2}\sum_{k=1}^N((x_k)_{t_1})^2-\sum_{k=1}^N e^{x_{k+1}-x_k},\\
L_2(x,x_{t_1},x_{t_2}) & = & \sum_{k=1}^N (x_k)_{t_1}(x_k)_{t_2}-\frac{1}{3}\sum_{k=1}^N((x_k)_{t_1})^3-\sum_{k=1}^N e^{x_{k+1}-x_k}\big((x_k)_{t_1}+(x_{k+1})_{t_1}\big).\qquad
\end{eqnarray}
Then one of the two equations (\ref{eq: offdiag}) is trivially satisfied, while the second one reads.
\begin{equation}\label{eq: Toda 2 xt}
(x_k)_{t_2}=((x_k)_{t_1})^2+e^{x_{k+1}-x_k}+e^{x_k-x_{k-1}}.
\end{equation}
Equation (\ref{eq: diag}) is satisfied identically. Finally, equations (\ref{eq: EL}) read:
\begin{eqnarray}
(x_k)_{t_1t_1} & = & e^{x_{k+1}-x_k}-e^{x_k-x_{k-1}},
  \label{eq: Toda 2 xtt1} \\
(x_k)_{t_1t_2} & = & e^{x_{k+1}-x_k}\big((x_k)_{t_1}+(x_{k+1})_{t_1}\big)-
                  e^{x_k-x_{k-1}}\big((x_{k-1})_{t_1}+(x_k)_{t_1}\big).
  \label{eq: Toda 2 xtt2}
\end{eqnarray}
Equations (\ref{eq: Toda 2 xt})--(\ref{eq: Toda 2 xtt2}) constitute the two-time Euler-Lagrange equations for the above Lagrangian 1-form. One shows by a direct computation that the compatibility condition of the latter two equations, $((x_k)_{t_1t_1})_{t_2}=((x_k)_{t_1t_2})_{t_1}$, is satisfied by virtue of equations (\ref{eq: Toda 2 xt}) and (\ref{eq: Toda 2 xtt1}). A well-posed initial value problem for equations (\ref{eq: Toda 2 xt})--(\ref{eq: Toda 2 xtt2}) can be obtained by prescribing $x$ and $x_{t_1}$ at {\em one point} $(t_1,t_2)=(0,0)$ of the two-time.

In order to arrive at more tractable conditions, we adopt the following definition.
\begin{definition}
We say that the multi-time Lagrangian 1-form $\cL$ is {\em Legendre transformable}, if the set of equations consisting of (\ref{eq: offdiag}) and
\begin{equation}\label{eq: diag with p}
    \frac{\partial L_1}{\partial v_1}(x,x_{t_1},\ldots,x_{t_m})
    %=\frac{\partial L_2}{\partial v_2}(x,x_{t_1},\ldots,x_{t_m})
    =\ldots=\frac{\partial L_m}{\partial v_m}(x,x_{t_1},\ldots,x_{t_m})=p
\end{equation}
can be solved for $x_{t_\alpha}$ in terms of $x,p$:
\begin{equation}\label{eq: for Legendre}
x_{t_\alpha}=x_{t_\alpha}(x,p),\quad \alpha=1,\ldots,m.
\end{equation}
\end{definition}
Clearly, for Legendre transformable 1-forms the right-hand sides of differential equations (\ref{eq: EL}) can be expressed through $x,p$, as well.

{\bf Example, continued.} In the example above, equations (\ref{eq: for Legendre}) take the form
\begin{eqnarray}
(x_k)_{t_1} & = & p_k,
   \label{eq: Toda 2 Ham x1}\\
(x_k)_{t_2} & = & p_k^2+e^{x_{k+1}-x_k}+e^{x_k-x_{k-1}},
   \label{eq: Toda 2 Ham x2}
\end{eqnarray}
while equations (\ref{eq: EL}) can be put as
\begin{eqnarray}
(p_k)_{t_1} & = & e^{x_{k+1}-x_k}-e^{x_k-x_{k-1}},
  \label{eq: Toda 2 Ham p1} \\
(p_k)_{t_2} & = & e^{x_{k+1}-x_k}(p_k+p_{k+1})-e^{x_k-x_{k-1}}(p_{k-1}+p_k).
  \label{eq: Toda 2 Ham p2}
\end{eqnarray}
One easily sees that equations (\ref{eq: Toda 2 Ham x1}), (\ref{eq: Toda 2 Ham p1}) are Hamiltonian with the Hamilton function
\[
H_1(x,p)=\frac{1}{2}\sum_{k=1}^N p_k^2+\sum_{k=1}^N e^{x_{k+1}-x_k},
\]
while equations (\ref{eq: Toda 2 Ham x2}), (\ref{eq: Toda 2 Ham p2}) are Hamiltonian with the Hamilton function
\[
H_2(x,p)=\frac{1}{3}\sum_{k=1}^N p_k^3+\sum_{k=1}^N e^{x_{k+1}-x_k}(p_{k+1}+p_k).
\]
These observations illustrate the following general statement.

\begin{theorem}\label{th: Ham}
Suppose that the multi-time Lagrangian 1-form $\cL$  is Legendre transformable. Define the Hamilton functions
\begin{equation}\label{eq: Legendre}
    H_\alpha(x,p)=\langle p,x_{t_\alpha}\rangle-L_\alpha(x,x_{t_1},\ldots,x_{t_m}),\quad \alpha=1,\ldots,m
\end{equation}
(it is understood that all $x_{t_\beta}$ on the right-hand side of the latter formula are expressed according to (\ref{eq: for Legendre})).
The functions $x,p:\bbR^m\to X$ satisfy Hamiltonian equations of motion:
\begin{equation}\label{eq: Ham eqs}
    \frac{\partial x}{\partial t_\alpha}=\frac{\partial H_\alpha}{\partial p},\quad
    \frac{\partial p}{\partial t_\alpha}=-\frac{\partial H_\alpha}{\partial x}.
\end{equation}
Hamiltonian flows with Hamilton functions $H_\alpha$ commute, so that their pairwise Poisson brackets are constant:
\begin{equation}\label{eq: almost commute}
    \{H_\alpha,H_\beta\}=h_{\alpha\beta}={\rm const}.
\end{equation}
\end{theorem}

\begin{theorem}\label{th: closed}
The following identities hold true on solutions of the Euler-Lagrange equations:
\begin{equation}\label{eq: almost closed}
    \frac{\partial}{\partial t_\alpha}L_\beta(x,x_{t_1},\ldots,x_{t_m})-
    \frac{\partial}{\partial t_\beta}L_\alpha(x,x_{t_1},\ldots,x_{t_m})=h_{\alpha\beta},
\end{equation}
with the constants $h_{\alpha\beta}$ from (\ref{eq: almost commute}). In particular, if the Hamilton functions $H_\alpha$ are in involution, $h_{\alpha\beta}=0$, then the form $\cL$ is closed on solutions of the Euler-Lagrange equations, so that the action functional $S_\Gamma$ in (\ref{eq: functional}) does not depend on the choice of the curve $\Gamma$ connecting two given points in $\bbR^m$.
\end{theorem}

Conversely, any system of commuting Hamiltonian flows admits a Lagrangian formulation with a 1-form satisfying conditions of Theorem \ref{th: EL}. A trivial version of such a formulation is obtained upon individual Legendre transformations of all functions $H_\alpha$, so that each $L_\alpha$ only depends on its own velocity: $L_\alpha=L_\alpha(x,x_{t_\alpha})$. However, this construction might be either impossible (if some of Legendre transformations are non-invertible, i.e., relations $x_{t_\alpha}=\partial H_\alpha/\partial p$ are not solvable for $p$ in terms of $x, x_{t_\alpha}$), or just impractical (if the above mentioned relations are solvable for $p$ but the resulting expressions are inconvenient, i.e., given by algebraic functions of high degree). The following theorem provides us with a convenient construction which works whenever just {\em one} of the Legendre transformations is easily invertible, which is usually the case in applications.

\begin{theorem}\label{th: constr}
For any system of $m$ commuting Hamiltonian flows with Hamilton functions $H_\alpha(x,p)$, $\alpha=1,\ldots,m$, one can find a multi-time Lagrangian 1-form satisfying conditions (and statements) of Theorem \ref{th: EL}. In particular, if the Hesse matrix $\partial^2H_1/\partial p^2$ is non-degenerate, so that the relations $x_{t_1}=\partial H_1/\partial p$ can be solved for $p=p(x,x_{t_1})$, then one can chose the components of $\cL$ as follows:
\begin{eqnarray}
L_1(x,x_{t_1}) & = & \langle p,x_{t_1}\rangle-H_1(x,p),\\
L_\alpha(x,x_{t_1},x_{t_\alpha}) & = & \langle p,x_{t_\alpha}\rangle-H_\alpha(x,p),\quad \alpha=2,\ldots,m
\end{eqnarray}
(with the understanding that $p$ on the right-hand sides of all these formulas is expressed through $x,x_{t_1}$).
\end{theorem}

%%%%%%%%%%%%%%%%%%%%%%%%%%%%%%%
%%%%%%%%%%%%%%%%%%%%%%%%%%%%%%%
\section{Proof of main theorems in the continuous time case}
\label{sect: cont proofs}
%%%%%%%%%%%%%%%%%%%%%%%%%%%%%%%
%%%%%%%%%%%%%%%%%%%%%%%%%%%%%%%

{\em Proof of Theorem \ref{th: EL}}. For a function $x:\bbR^m\to X$ and for a fixed curve $\Gamma:[0,1]\to\bbR^m$, we denote by $x_\Gamma=x\circ\Gamma:[0,1]\to X$ the corresponding curve in $X$. A variation of the action functional $S_\Gamma$ due to an arbitrary variation $\delta x:\bbR^m\to X$ of the function $x$ can be written as
\begin{equation}\label{eq: delta S}
\delta S_\Gamma=\int_0^1\sum_{\alpha=1}^m\left(\left\langle\frac{\partial L_\alpha}{\partial x},\delta x\right\rangle+\sum_{\beta=1}^m\left\langle\frac{\partial L_\alpha}{\partial v_\beta},\delta x_{t_\beta}\right\rangle\right)\Gamma_\alpha'(s)ds
\end{equation}
Here (and in similar context below) it is understood that all terms in the parentheses are evaluated at $\Gamma(s)$. The partial derivatives of $\delta x$, when restricted to $\Gamma$, satisfy the following two conditions:
\begin{itemize}
\item The derivative of $\delta x$ in the tangential direction of $\Gamma$ is given by
\begin{equation}\label{eq: tang}
    \sum_{\alpha=1}^m \Gamma_\alpha'(s)\delta x_{t_\alpha}=(\delta x_\Gamma)',
\end{equation}
where $\delta x_\Gamma:[0,1]\to X$ is an arbitrary variation of the function $x_\Gamma$.
\item The derivatives of $\delta x$ in the normal directions to the curve $\Gamma$ are arbitrary, i.e., are independent of $\delta x_\Gamma$. More precisely, if $n_1(s),\ldots,n_{m-1}(s)$ is an arbitrary basis of $(\Gamma'(s))^\perp$, the normal space to the curve $\Gamma$ at the point $\Gamma(s)$, then
\begin{equation}\label{eq: normal}
   \sum_{\beta=1}^m n_{\gamma\beta} \delta x_{t_\beta}=\delta y_\gamma,\quad \gamma=1,\ldots,m-1,
\end{equation}
    where $n_{\gamma\beta}$ are the components of the vector $n_\gamma$, and $\delta y_\gamma$ are $m-1$ arbitrary functions on $\Gamma$.
\end{itemize}
The solution of system (\ref{eq: tang}), (\ref{eq: normal}) is most easily found, if we assume vectors $n_1(s),\ldots,n_{m-1}(s)$ to be orthonormal, and is then given by
\begin{equation}\label{eq: derivatives}
    \delta x_{t_\alpha}=\frac{\Gamma_\alpha'(s)}{\|\Gamma'(s)\|^2}(\delta x_\Gamma)'+\sum_{\gamma=1}^{m-1}n_{\gamma\alpha}(s)\delta y_\gamma.
\end{equation}
Now we are in a position to find necessary and sufficient conditions for $x$ to ensure $\delta S_\Gamma=0$. We have:
\begin{eqnarray}\label{eq: delta S 2}
\delta S_\Gamma & = & \int_0^1\left\langle\sum_{\alpha=1}^m\Gamma_\alpha'(s)\frac{\partial L_\alpha}{\partial x},\delta x_\Gamma\right\rangle ds+\int_0^1 \left\langle\sum_{\alpha=1}^m\sum_{\beta=1}^m
\frac{\Gamma_\alpha'(s)\Gamma_\beta'(s)}{\|\Gamma'(s)\|^2}\frac{\partial L_\alpha}{\partial v_\beta},(\delta x_\Gamma)' \right\rangle ds\nonumber\\
&& +\sum_{\gamma=1}^{m-1}\int_0^1\left\langle \sum_{\alpha=1}^m\sum_{\beta=1}^m
\Gamma_\alpha'(s)\frac{\partial L_\alpha}{\partial v_\beta}n_{\gamma\beta}(s),\delta y_\gamma\right\rangle ds.
\end{eqnarray}
By the usual integration by part argument in conjunction with the fact that $\delta x_\Gamma$ and $\delta y_\gamma$ ($\gamma=1,\ldots,m-1$) are arbitrary independent functions on $\Gamma$, we come to the following necessary and sufficient conditions of $\delta S_\Gamma=0$:
\begin{equation}\label{eq: EL prelim 1}
    \sum_{\alpha=1}^m\Gamma_\alpha'(s)\frac{\partial L_\alpha}{\partial x}-
    \frac{d}{ds}\sum_{\alpha=1}^m\sum_{\beta=1}^m
\frac{\Gamma_\alpha'(s)\Gamma_\beta'(s)}{\|\Gamma'(s)\|^2}\frac{\partial L_\alpha}{\partial v_\beta}=0,
\end{equation}
\begin{equation}\label{eq: EL prelim 2}
    \sum_{\alpha=1}^m\sum_{\beta=1}^m
\Gamma_\alpha'(s)\frac{\partial L_\alpha}{\partial v_\beta}n_{\gamma\beta}(s)=0, \quad \gamma=1,\ldots,m-1.
\end{equation}
The arguments of the functions $\partial L_\alpha/\partial x$, $\partial L_\alpha/\partial v_\beta$ in these formulas are $(x,x_{t_1},\ldots,x_{t_m})$ for $x:\bbR^m\to X$, evaluated at $\Gamma(s)$. Once again, equations (\ref{eq: EL prelim 1}), (\ref{eq: EL prelim 2}) characterize functions $x$ delivering a critical point for the functional $S_\Gamma$ for a fixed curve $\Gamma$.

Since the curve $\Gamma$ is arbitrary, we can further argue as follows. Equation (\ref{eq: EL prelim 2}) says that
\[
  \sum_{\alpha=1}^m\sum_{\beta=1}^m
\Gamma_\alpha'(s)\frac{\partial L_\alpha}{\partial v_\beta}n_{\beta}(s)=0
\]
for an arbitrary pair of orthogonal vectors $\Gamma'(s)$ and $n(s)$ at an arbitrary point $\Gamma(s)\in\bbR^m$. A necessary and sufficient condition for this is given by (\ref{eq: offdiag}), (\ref{eq: diag}). Substituting the latter into equation (\ref{eq: EL prelim 1}) results in
\[
\sum_{\alpha=1}^m\Gamma_\alpha'(s)\frac{\partial L_\alpha}{\partial x}-\frac{dp}{ds}=0,
\]
which is equivalent to (\ref{eq: EL}). \endpf
\medskip

{\em Proof of Theorem \ref{th: Ham}.} This is a standard computation, not much different from the one in the 1-time situation.
We compute:
\[
\frac{\partial H_\alpha}{\partial p}=
x_{t_\alpha}+\left\langle p,\frac{\partial x_{t_\alpha}}{\partial p}\right\rangle
-\sum_{\beta=1}^m\left\langle \frac{\partial L_\alpha}{\partial v_\beta},\frac{\partial x_{t_\beta}}{\partial p}\right\rangle=x_{t_\alpha}.
\]
Indeed, due to (\ref{eq: offdiag}), the sum reduces to one term with $\beta=\alpha$, which cancels with the second summand on the right-hand side, due to (\ref{eq: diag}). This proves the first equation of motion in (\ref{eq: Ham eqs}).

Literally the same argument justifies the following computation:
\[
\frac{\partial H_\alpha}{\partial x}=
\left\langle p,\frac{\partial x_{t_\alpha}}{\partial x}\right\rangle
-\frac{\partial L_\alpha}{\partial x}
-\sum_{\beta=1}^m\left\langle \frac{\partial L_\alpha}{\partial v_\beta},\frac{\partial x_{t_\beta}}{\partial x}\right\rangle=-\frac{\partial L_\alpha}{\partial x}.
\]
This proves the second equation of motion in (\ref{eq: Ham eqs}), by virtue of (\ref{eq: EL}):
\[
\frac{\partial p}{\partial t_\alpha}=\frac{\partial L_\alpha}{\partial x}=
-\frac{\partial H_\alpha}{\partial x}.
\]
Thus, Hamiltonian equations of motion are verified. Commutativity of flows is a mere reformulation of the existence of functions $x,p:\bbR^m\to X$ which solve all equations (\ref{eq: Ham eqs}) simultaneously. \endpf
\medskip

{\em Proof of Theorem \ref{th: closed}.} A very similar computation:
\[
\frac{\partial}{\partial t_\beta}L_\alpha(x,x_{t_1},\ldots,x_{t_m})=
\left\langle\frac{\partial L_\alpha}{\partial x},\frac{\partial x}{\partial t_\beta}\right\rangle+
\sum_{\gamma=1}^m \left\langle\frac{\partial L_\alpha}{\partial x_{t_\gamma}},\frac{\partial x_{t_\gamma}}{\partial t_\beta}\right\rangle.
\]
Upon using equations of motion (\ref{eq: EL}), (\ref{eq: offdiag}), (\ref{eq: diag}), we come to
\[
\frac{\partial}{\partial t_\beta}L_\alpha(x,x_{t_1},\ldots,x_{t_m})=
\left\langle\frac{\partial p}{\partial t_\alpha},
\frac{\partial x}{\partial t_\beta}\right\rangle+
\left\langle p,\frac{\partial^2 x}{\partial t_\alpha\partial t_\beta}\right\rangle.
\]
This yields
\[
\frac{\partial}{\partial t_\beta}L_\alpha(x,x_{t_1},\ldots,x_{t_m})-
\frac{\partial}{\partial t_\alpha}L_\beta(x,x_{t_1},\ldots,x_{t_m})=
\left\langle\frac{\partial p}{\partial t_\alpha},
\frac{\partial x}{\partial t_\beta}\right\rangle-
\left\langle\frac{\partial p}{\partial t_\beta},
\frac{\partial x}{\partial t_\alpha}\right\rangle,
\]
and, by virtue of Hamiltonian equations of motion (\ref{eq: Ham eqs}),
\[
\frac{\partial}{\partial t_\beta}L_\alpha(x,x_{t_1},\ldots,x_{t_m})-
\frac{\partial}{\partial t_\alpha}L_\beta(x,x_{t_1},\ldots,x_{t_m})=
-\left\langle\frac{\partial H_\alpha}{\partial x},
\frac{\partial H_\beta}{\partial p}\right\rangle+
\left\langle\frac{\partial H_\beta}{\partial x},
\frac{\partial H_\alpha}{\partial p}\right\rangle,
\]
which is the desired result. \endpf
\medskip

{\em Proof of Theorem \ref{th: constr}}. A direct verification of conditions of Theorem \ref{th: EL}. \endpf

%%%%%%%%%%%%%%%%%%%%%%%%%%%%%%%
%%%%%%%%%%%%%%%%%%%%%%%%%%%%%%%
\section{Basic construction in the discrete time case}
\label{sect: discr results}
%%%%%%%%%%%%%%%%%%%%%%%%%%%%%%%
%%%%%%%%%%%%%%%%%%%%%%%%%%%%%%%

In this section, we consider a sort of a discretization of the previous construction. We are looking for functions $x:\bbZ^m\to X=\bbR^N$ which deliver critical points for the following variational problem. Let $\Gamma$ be an arbitrary discrete curve (path) which is a concatenation of a sequence of directed edges $\mathfrak{e}_k$ in $\bbZ^m$ such that the endpoint of any $\mathfrak{e}_k$ is the beginning of $\mathfrak{e}_{k+1}$, we set
\begin{equation}\label{eq: discr action}
    S_\Gamma=\sum_{k\in\bbZ}\cL(\mathfrak{e}_k).
\end{equation}
Here $\cL$ is a discrete 1-form, i.e., a skew-symmetric function on directed edges of the regular square lattice $\bbZ^m$, defined as follows:
\[
\cL(n,n+e_\alpha)=\Lambda_\alpha(x,x_\alpha)\quad\Leftrightarrow\quad \cL(n,n-e_\alpha)=-\Lambda_\alpha(x_{-\alpha},x).
\]
Here $\Lambda_\alpha:X\times X\to\bbR$ are local Lagrangian functions corresponding to the edges of the $\alpha$-th coordinate direction, and the following abbreviations are used: $x$ for $x(n)$ at a generic point $n\in\bbZ^m$, and then
\begin{equation}\label{eq: shifts}
    x_\alpha=x(n+e_\alpha),\quad x_{-\alpha}=x(n-e_\alpha), \quad \alpha=1,\ldots,m.
\end{equation}
Here $e_\alpha$ stands for the unit vector of the $\alpha$-th coordinate direction.

\begin{theorem}\label{th: dEL}
The function $x:\bbZ^m\to X$ delivers a critical point for the functional $S_\Gamma$ for any discrete curve $\Gamma$, if and only if the following conditions (which we call {\em multi-time discrete Euler-Lagrange equations}) are satisfied:
\begin{eqnarray}
\frac{\partial\Lambda_\alpha(x_{-\alpha},x)}{\partial x}+
\frac{\partial\Lambda_\beta(x,x_{\beta})}{\partial x} & = & 0,
\label{eq: dEL -alpha beta}\\
\frac{\partial\Lambda_\alpha(x_{-\alpha},x)}{\partial x}-
\frac{\partial\Lambda_\beta(x_{-\beta},x)}{\partial x} & = & 0,
\label{eq: dEL -alpha -beta}\\
\frac{\partial\Lambda_\alpha(x,x_{\alpha})}{\partial x}-
\frac{\partial\Lambda_\beta(x,x_{\beta})}{\partial x} & = & 0.
\label{eq: dEL alpha beta}
\end{eqnarray}
In other words, there exists a function $p:\bbZ^m\to X$ satisfying all the relations
\begin{eqnarray}
    p & = & \frac{\partial\Lambda_\alpha(x,x_\alpha)}{\partial x},\quad \alpha=1,\ldots,m,
    \label{eq: discr p 1}\\
    p  & = & -\frac{\partial\Lambda_\alpha(x_{-\alpha},x)}{\partial x},\quad \alpha=1,\ldots,m.
     \label{eq: discr p 2}
\end{eqnarray}
\end{theorem}

Continuing analogy with the continuous time case, we introduce the following definition.
\begin{definition}\label{def: discr Legendre}
We say that the multi-time discrete Lagrangian 1-form $\cL$ is {\em Legendre transformable}, if all the equations (\ref{eq: discr p 1})
can be solved for $x_\alpha$ in terms of $x,p$.
\end{definition}
\begin{theorem}\label{th: commuting maps}
Suppose that the multi-time discrete Lagrangian 1-form is Legendre transformable. Then equations \begin{equation}\label{eq: single Lagr map}
    p=\frac{\partial\Lambda_\alpha(x,x_{\alpha})}{\partial x},\quad
    p_\alpha=-\frac{\partial\Lambda_\alpha(x,x_{\alpha})}{\partial x_\alpha},
\end{equation}
determine a symplectic map $F_\alpha:(x,p)\mapsto(x_\alpha,p_\alpha)$. These maps for different $\alpha$ commute:
\begin{equation}\label{eq: commute}
F_\alpha\circ F_\beta=F_\beta\circ F_\alpha.
\end{equation}
\end{theorem}
\begin{theorem}\label{th: discr almost closed}
The following identities hold true on solutions of the discrete Euler-Lagrange equations:
\begin{equation}\label{eq: discr almost closed}
    \Lambda_\alpha(x,x_\alpha)+\Lambda_\beta(x_\alpha,x_{\alpha\beta})-
    \Lambda_\alpha(x_\beta,x_{\alpha\beta})-\Lambda_\beta(x,x_\beta)=\ell_{\alpha\beta}=
    {\rm const}.
\end{equation}
In particular, if all these constants $\ell_{\alpha\beta}$ vanish, then the discrete 1-form $\cL$ is closed on solutions of the Euler-Lagrange equations, so that the action functional $S_\Gamma$ does not depend on the choice of the curve $\Gamma$ connecting two given points in $\bbZ^m$.
\end{theorem}

In the full generality, the above discrete theory does not lead to any statements about integrability of the maps $F_\alpha$. However, such statements become possible for an important class of examples, namely, for {\em B\"acklund transformations}. We understand B\"acklund transformations as a one-parameter family of commuting symplectic maps. Thus, our point of view is in a sense opposite to that of Kuznetsov and Sklyanin in \cite{KS}. While the primary feature of B\"acklund transformations for them was the existence of a common complete set of integrals in involution coming from a common Lax matrix (and commutativity was considered as a consequence of this property by virtue of the discrete Liouville-Arnold theorem), we propose to put an emphasis on the commutativity property.
Actually, we even do not pre-suppose the existence of the Lax representation for the B\"acklund transformations, but rather derive it from the maps themselves.

Thus, we consider the family of maps $F_\lambda:\bbR^{2N}\to\bbR^{2N}$, $(x,p)\mapsto(\wx,\wip)$, given by equations of the type (\ref{eq: single Lagr map}):
\begin{equation}\label{eq: BT1}
F_\lambda:\
p=\frac{\partial \Lambda(x,\wx;\lambda)}{\partial x},\quad
\wip=-\frac{\partial \Lambda(x,\wx;\lambda)}{\partial \wx}.
\end{equation}
When considering a second such map, say $F_\mu$, we will denote its action by a hat:
\begin{equation}\label{eq: BT2}
F_\mu:\
p=\frac{\partial \Lambda(x,\whx;\mu)}{\partial x},\quad
\whp=-\frac{\partial \Lambda(x,\whx;\mu)}{\partial \whx}.
\end{equation}
Assuming that $F_\lambda$, $F_\mu$ commute for any $\lambda$ and $\mu$, we write the result of Theorem \ref{th: discr almost closed} as
\begin{equation}\label{eq: BT closure}
    \Lambda(x,\wx;\lambda)+\Lambda(\wx,\widehat{\wx};\mu)-
    \Lambda(x,\whx;\mu)-\Lambda(\whx,\widehat{\wx};\lambda)=\ell(\lambda,\mu).
\end{equation}
Clearly, the possible dependence of this constant on the parameters $\lambda$, $\mu$ is skew-symmetric: $\ell(\lambda,\mu)=-\ell(\mu,\lambda)$.
\begin{theorem}\label{th: spectrality}
For a family of B\"acklund transformations, the discrete 1-form $\mathcal L$ is closed on solutions of the multi-time Euler Lagrange equations, i.e., $\ell(\lambda,\mu)=0$, if and only if $\partial \Lambda(x,\wx;\lambda)/\partial \lambda$ is a common integral of motion for all $F_\mu$.
\end{theorem}
The latter property is a re-formulation of the {\em spectrality property} of B\"acklund transformations introduced by Kuznetsov and Sklyanin \cite{KS} but stripped of its mystical flavor by not mentioning the additional structure of the zero-curvature, or Lax, representation.

Turning the subject of the Lax representation, we would like to pursue the one-dimensional analogue of the basic idea of consistency as integrability \cite{BS1}, \cite{N}. According to this idea, a discrete multi-dimensionally consistent two-dimensional system serves as its own zero curvature representation. Our message concerning the one-dimensional case is:
\begin{itemize}
\item[] {\em Given a one-parameter family of commuting symplectic maps, one can derive a Lax (or zero-curvature) representation with a spectral parameter from the maps themselves.}
\end{itemize}
This can be considered as a further step to de-mystifying integrable systems. Indeed, commutativity is a simple property allowing for a direct (may be, computer aided) check, while Lax representations are transcendental objects whose mere existence has always been considered as the most mysterious feature of integrable systems.

We will not give a general derivation, but instead present it in the particular case of B\"acklund transformations for the Toda lattice described in Section \ref{sect: BT Toda}. An extensive body of concrete results for further families of B\"acklund transformations will be presented elsewhere.

%%%%%%%%%%%%%%%%%%%%%%%%%%%%%%%
%%%%%%%%%%%%%%%%%%%%%%%%%%%%%%%
\section{Proofs of main theorems in the discrete time case}
\label{sect: discr proofs}
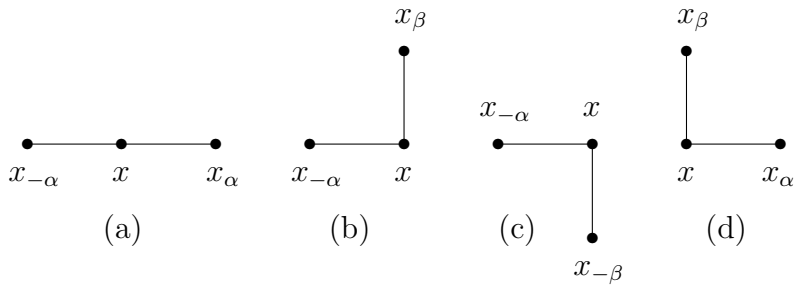
\begin{figure}[htbp]
    \setlength{\unitlength}{0.06em}
\begin{picture}(400,150)(-50,-70)
  \put(-50, 0){\circle*{6}} \put(0,0){\circle*{6}}\put(50,0){\circle*{6}}
  \put(-50,  0){\line(1,0){100}}
  \put(-60,-20){$x_{-\alpha}$}
  \put(-5,-20){$x$}
  \put(45,-20){$x_\alpha$}
  \put(-10,-50){(a)}

  \put(100,0){\circle*{6}} \put(150,0){\circle*{6}}\put(150,50){\circle*{6}}
  \put(100,0){\line(1,0){50}}\put(150,0){\line(0,1){50}}
  \put(90,-20){$x_{-\alpha}$}
  \put(145,-20){$x$}
  \put(145,65){$x_\beta$}
  \put(110,-50){(b)}

  \put(200,0){\circle*{6}} \put(250,0){\circle*{6}}\put(250,-50){\circle*{6}}
  \put(200,0){\line(1,0){50}}\put(250,-50){\line(0,1){50}}
  \put(190,15){$x_{-\alpha}$}
  \put(245,15){$x$}
  \put(240,-70){$x_{-\beta}$}
  \put(200,-50){(c)}

  \put(300,0){\circle*{6}} \put(350,0){\circle*{6}}\put(300,50){\circle*{6}}
  \put(300,0){\line(1,0){50}}\put(300,0){\line(0,1){50}}
  \put(295,-20){$x$}
  \put(340,-20){$x_\alpha$}
  \put(295,65){$x_\beta$}
  \put(310,-50){(d)}
\end{picture}
\caption{Four type of vertices of a discrete curve. Cases (a) and (b): two equally (positively or negatively) directed edges meet at $n$, of one and of two different coordinate directions, repectively. Case (c): a positively directed adges followed by a negatively directed edge. Case (d): a negatively directed edge followed by a positively directed edge.}
\label{Fig: corners}
\end{figure}

%%%%%%%%%%%%%%%%%%%%%%%%%%%%%%%
%%%%%%%%%%%%%%%%%%%%%%%%%%%%%%%
{\em Proof of Theorem \ref{th: dEL}}. At any point of a discrete curve there meet two directed edges. It is easy to understand that, up to an overall change of orientation of the curve and up to permutations of indices, only four possibilities exist, depicted on Figure \ref{Fig: corners}.
%--------------------------------------------------------------------------

Correspondingly, there are four types of the Euler-Lagrange equations. The first one is for a straight piece of a discrete curve, as on Fig.~\ref{Fig: corners}, (a):
\begin{equation}\label{eq: dEL -alpha alpha}
\frac{\partial\Lambda_\alpha(x_{-\alpha},x)}{\partial x}+
\frac{\partial\Lambda_\alpha(x,x_{\alpha})}{\partial x}=0,
\end{equation}
while the other three correspond to the corner type pieces of a discrete curve, as on Fig.~\ref{Fig: corners}, (b)--(d), and are given in (\ref{eq: dEL -alpha beta})--(\ref{eq: dEL alpha beta}). It is important to observe that these equations are not independent, for instance (\ref{eq: dEL -alpha alpha}) follows from (\ref{eq: dEL -alpha beta}) and (\ref{eq: dEL alpha beta}). Existence of the function $p:\bbZ^m\to X$ is a direct corollary of equations (\ref{eq: dEL -alpha beta})--(\ref{eq: dEL alpha beta}). \endpf
\medskip

{\em Proof of Theorem \ref{th: commuting maps}}. The symplectic property of the maps $F_\alpha$ was proven by Moser-Veselov \cite{MV}. The commutativity of maps $F_\alpha$ is a mere restatement of the existence of functions $x,p:\bbZ^m\to X$ which solve all equations (\ref{eq: discr p 1}), (\ref{eq: discr p 2}) simultaneously. We remark that a function $(x,p):\bbZ^m\to\bbR^{2N}$ such that $(x(n+e_\alpha),p(n+e_\alpha))=F_\alpha(x(n),p(n))$ is defined by a single initial value $(x(0),p(0))$. \endpf
\medskip

{\em Proof of Theorem \ref{th: discr almost closed}}. Denote the left-hand side of equation (\ref{eq: discr almost closed}) by $\Omega$. It is a function on the manifold of solutions of discrete Euler-Lagrange equations, which is, according to the last remark in the previous proof, $(2N)$-dimensional, as it can parametrized by $(x,p)$, or by $(x,x_\alpha)$ with some fixed $\alpha$. It is enough to prove that $\partial \Omega/\partial x=0$ and $\partial \Omega/\partial x_\alpha=0$. We prove a stronger statement: if one considers $\Omega$ as a function on the $(4N)$-dimensional manifold, parametrized by $x$, $x_\alpha$, $x_\beta$, and $x_{\alpha\beta}$, then the gradient of this function vanishes on the $(2N)$-dimensional submanifold of solutions of discrete Euler-Lagrange equations. But this is obvious, since vanishing of the partial derivatives of $\Omega$ with respect to its 4 vector-valued arguments is nothing but the discrete Euler-Lagrange equations at the corresponding points. \endpf
\medskip

{\em Proof of Theorem \ref{th: spectrality}}. Due to skew-symmetry, $\ell(\lambda,\mu)=0$ is equivalent to $\partial \ell(\lambda,\mu)/\partial \lambda=0$, that is, to
\[
 \frac{\partial\Lambda(x,\wx;\lambda)}{\partial \lambda}
 -\frac{\partial\Lambda(\whx,\widehat{\wx};\lambda)}{\partial \lambda}=0
\]
(see equation (\ref{eq: BT closure})). This is equivalent to saying that $\partial \Lambda(x,\wx;\lambda)/\partial \lambda$ is an integral of motion for $F_\mu$. \endpf

%%%%%%%%%%%%%%%%%%%%%%%%%%%%%%%
%%%%%%%%%%%%%%%%%%%%%%%%%%%%%%%
\section{Example: B\"acklund transformations for Toda lattice}
\label{sect: BT Toda}
%%%%%%%%%%%%%%%%%%%%%%%%%%%%%%%
%%%%%%%%%%%%%%%%%%%%%%%%%%%%%%%

Here we illustrate the main constructions of the discrete time case by the well-known example of B\"acklund transformations for the Toda lattice \cite{TW}, \cite{KS}, also known as discrete time Toda lattice \cite{S}. From this one-parameter family of commuting symplectic maps, we consider just two (for notational simplicity only; considering any $m$ would go along literally the same lines). Our maps $F_\lambda, F_\mu:\bbR^{2N}\to\bbR^{2N}$ are given by equations of the type (\ref{eq: single Lagr map}):
\begin{equation}\label{eq: BT1 Toda}
F_\lambda:\
p_k=\dfrac{1}{\lambda}\left(e^{\wx_k-x_k}-1\right)+\lambda e^{x_k-\wx_{k-1}},\quad
\wip_k=\dfrac{1}{\lambda}\left(e^{\wx_k-x_k}-1\right)+\lambda e^{x_{k+1}-\wx_k},
\end{equation}
\begin{equation}\label{eq: BT2 Toda}
F_\mu:\
p_k=\dfrac{1}{\mu}\left(e^{\whx_k-x_k}-1\right)+\mu e^{x_k-\whx_{k-1}},\quad
\whp_k=\dfrac{1}{\mu}\left(e^{\whx_k-x_k}-1\right)+\mu e^{x_{k+1}-\whx_k}.
\end{equation}
The corresponding Lagrangians are given by
\begin{eqnarray}
\Lambda(x,\wx;\lambda) & = & \frac{1}{\lambda}\sum_{k=1}^N\left(e^{\wx_k-x_k}-1-(\wx_k-x_k)\right)-
    \lambda\sum_{k=1}^N e^{x_{k+1}-\wx_k}, \label{eq: BT1 Lagr}\\
\Lambda(x,\whx;\mu) & = & \frac{1}{\mu}\sum_{k=1}^N\left(e^{\whx_k-x_k}-1-(\whx_k-x_k)\right)-
    \mu\sum_{k=1}^N e^{x_{k+1}-\whx_k}. \label{eq: BT2 Lagr}
\end{eqnarray}
These maps are considered under open-end boundary conditions ($x_0=\infty$, $x_{N+1}=-\infty$) or under periodic boundary conditions (all indices are taken ${\rm mod}\, N$, so that $x_0=x_N$, $x_{N+1}=x_1$).
The corresponding discrete Lagrangian 1-form is Legendre transformable. To see this, consider the map $F_\lambda$. In the open-end case, the first equations in (\ref{eq: BT1 Toda}) are uniquely solved for $\wx_1$, $\wx_2$, $\ldots$, $\wx_N$ (in this order), to give
\[
e^{\wx_1-x_1}=1+\lambda p_1,\quad
e^{\wx_2-x_2}=1+\lambda p_2-\frac{\lambda^2e^{x_2-x_1}}{1+\lambda p_1},\quad\cdots\quad,
\]
\begin{equation}\label{eq: BT Toda expl}
e^{\wx_N-x_N}=1+\lambda p_N-\frac{\lambda^2e^{x_N-x_{N-1}}}{1+\lambda p_{N-1}-
\dfrac{\lambda^2e^{x_{N-1}-x_{N-2}}}{1+\lambda p_{N-2}-\;
\raisebox{-3mm}{$\ddots$}
\raisebox{-4.5mm}{$\;-\dfrac{\lambda^2 e^{x_2-x_1}}{1+\lambda p_1}$}}}.
\end{equation}
In the periodic case $e^{\wx_k-x_k}$ can be expressed as analogous infinite periodic continued fractions and are, therefore, {\it double-valued} functions of $(x,p)$.

It is well known that the maps $F_\lambda$, $F_\mu$ commute, see, e.g., \cite{TW}, \cite{KS}, \cite{S} (however, the very notion of commutativity for double-valued maps, relevant for the periodic case, requires for a more careful discussion, which we will provide in a separate publication). The most direct way towards the proof of commutativity relies on the following superposition formula:
\begin{equation}\label{eq: BT superp}
    e^{\widehat{\widetilde{x}}_k-\wx_k-\whx_k+x_{k+1}}=
    \dfrac{\lambda e^{\whx_{k+1}}-\mu e^{\wx_{k+1}}}{\lambda e^{\whx_{k}}-\mu e^{\wx_{k}}},
\end{equation}
which is also equivalent to either of the two formulas
\begin{eqnarray}
  \dfrac{1}{\lambda}e^{\wx_{k+1}-x_{k+1}}-\dfrac{1}{\mu}e^{\whx_{k+1}-x_{k+1}}
  -\dfrac{1}{\lambda} e^{\widehat{\widetilde{x}}_k-\whx_k}
  +\dfrac{1}{\mu} e^{\widehat{\widetilde{x}}_k-\wx_k} &=& 0,
  \label{eq: BT superp 1}\\
  \lambda e^{x_{k+1}-\wx_k}-\mu e^{x_{k+1}-\whx_k}
  -\lambda e^{\whx_{k+1}-\widehat{\widetilde{x}}_k}
  +\mu e^{\wx_{k+1}-\widehat{\widetilde{x}}_k} & = & 0.
  \label{eq: BT superp 2}
\end{eqnarray}
Thus, for any initial point $(x(0),p(0))$ maps $F_\lambda$ and $F_\mu$ determine a function $x:\bbZ^2\to\bbR^N$ which, according to Theorem \ref{th: dEL}, delivers a critical point to the action defined by an arbitrary curve $\Gamma$ in $\bbZ^2$ and by the Lagrangian 1-form $\mathcal L$ with the Lagrangians $\Lambda(x,\wx;\lambda)$ and $\Lambda(x,\whx;\mu)$ on the edges parallel to the first (resp. second) coordinate direction and satisfies the following discrete Euler-Lagrange equations:
\begin{eqnarray}\label{eq: BT EL}
 p_k & = & \dfrac{1}{\lambda}\left(e^{\wx_k-x_k}-1\right)+\lambda e^{x_k-\wx_{k-1}} \ = \
 \dfrac{1}{\lambda}\left(e^{x_k-\undertilde{x}_k}-1\right)+\lambda e^{\undertilde{x}_{k+1}-x_k} \\
 & = &  \dfrac{1}{\mu}\left(e^{\whx_k-x_k}-1\right)+\mu e^{x_k-\whx_{k-1}} \ = \
 \dfrac{1}{\mu}\left(e^{x_k-\underhat{x}_k}-1\right)+\mu e^{\underhat{x}_{k+1}-x_k}.
\end{eqnarray}
Moreover, we can prove that the Lagrangian 1-form $\mathcal L$ is closed on solutions. For this, we first show that this property, i.e., $\ell(\lambda,\mu)=0$ for $\ell(\lambda,\mu)$ from (\ref{eq: BT closure}), is equivalent to
\begin{equation}\label{eq: for closeness}
\sum_{k=1}^N(\widehat{\widetilde{x}}_k-\wx_k-\whx_k+x_k)=0\quad \Leftrightarrow\quad
\prod_{k=1}^N e^{\widehat{\widetilde{x}}_k-\wx_k-\whx_k+x_k}=1.
\end{equation}
For this aim, we observe that, by virtue of the superposition formulas (\ref{eq: BT superp 1}), (\ref{eq: BT superp 2}), most of the terms on left-hand side of (\ref{eq: BT closure}) cancel, leaving us with
\[
\ell(\lambda,\mu)=\left(\frac{1}{\lambda}-\frac{1}{\mu}\right)
\sum_{k=1}^N(\widehat{\widetilde{x}}_k-\wx_k-\whx_k+x_k).
\]
Alternatively, one can refer to Theorem \ref{th: spectrality}. Indeed, one easily computes:
\[
\frac{\partial \Lambda(x,\wx;\lambda)}{\partial \lambda}=-\frac{1}{\lambda}\sum_{k=1}^N p_k+\frac{1}{\lambda^2}\sum_{k=1}^N(\wx_k-x_k),
\]
the first sum on the right-hand side being an obvious integral of motion. Now the desired result (\ref{eq: for closeness}) follows in the periodic case by multiplying equations (\ref{eq: BT superp}) for $1\le k\le N$, in the open-end case equation (\ref{eq: BT superp}) holds true for $1\le k\le N-1$ and has to be supplemented by the boundary counterparts
\begin{equation}\label{eq: BT superp open boundary}
 e^{x_{1}}=\dfrac{\lambda e^{\whx_1}-\mu e^{\wx_1}}{\lambda-\mu}, \quad
 e^{\widehat{\widetilde{x}}_N-\wx_N-\whx_N}=\dfrac{\lambda-\mu}{\lambda e^{\whx_N}-\mu e^{\wx_N}}.
\end{equation}
Thus, spectrality of B\"acklund transformations for the Toda lattice is a direct consequence of the superposition formula (\ref{eq: BT superp}).
\medskip

Finally, we turn to the topic of Lax, or zero-curvature representations for B\"acklund transformations. We consider the map $F_\lambda$ as the basic object (it represents actually the whole family of maps, since the parameter $\lambda$ is arbitrary), and will produce the zero curvature representation from the map $F_\mu$, with $\mu$ playing the role of the spectral parameter. For this aim, we will extract the ``wave functions'' $\psi_k$ from the variables $\whx_k$. Indeed, setting
\[
e^{\whx_k-x_k}=\mu \frac{\psi_{k+1}}{\psi_k}
\]
(which defines $\psi_k$ up to a common factor), we re-write the first equation in (\ref{eq: BT2 Toda}) as
\[
\psi_{k+1}=\left(\frac{1}{\mu}+p_k\right)\psi_k-e^{x_k-x_{k-1}}\psi_{k-1}.
\]
This is nothing but the first (spectral) part of the inverse spectral formulation of the Toda hierarchy.

This first part of derivation was known from the early days of the soliton theory. For instance, it can be found in the paper \cite{TW} by Wadati and Toda from 1975. They treated B\"acklund transformations as symplectic maps commuting with the respective continuous time flows, and correspondingly demonstrated how to derive a Lax representation for a continuous flow from a B\"acklund transformation. This amounts to deriving a linear differential equation describing the temporal evolution of the wave function $\psi_k$. Thus, we depart from their derivation only now, by considering a commuting map instead of a commuting flow, which will lead to a difference equation for the discrete time evolution of the wave function $\psi_k$. It is amazing that this seemingly simple step (leading to a great conceptual simplification) took about 40 years to be done.
\smallskip

Now, we turn to deriving a consequence of the commutativity of $F_\lambda$ and $F_\mu$, expressed via the superposition formula (\ref{eq: BT superp}). It can be re-written as
\[
\mu\frac{\widetilde{\psi}_{k+1}}{\widetilde{\psi}_k}=e^{\widehat{\widetilde{x}}_k-\wx_k}=
\frac{\lambda e^{\whx_{k+1}-x_{k+1}}-\mu e^{\wx_{k+1}-x_{k+1}}}
{\lambda-\mu e^{\wx_k-x_k}e^{x_k-\whx_k}}=
\mu\frac{\lambda\psi_{k+2}-e^{\wx_{k+1}-x_{k+1}}\psi_{k+1}}
{\lambda\psi_{k+1}- e^{\wx_k-x_k}\psi_k},
\]
so that the second (evolutionary) part of the inverse spectral formulation is obtained:
\[
\widetilde{\psi}_k=-\lambda \psi_{k+1}+ e^{\wx_k-x_k}\psi_k.
\]
We get a standard zero-curvature representation of the B\"acklund transformation for the Toda lattice (see, e.g., \cite{KS}, \cite{S}) upon introducing an auxiliary wave function $\phi_k=-e^{-x_{k-1}}\psi_{k-1}$. Then we arrive at
\begin{equation}\label{eq: BT zcr 1}
\begin{pmatrix}\psi_{k+1} \\ \phi_{k+1}\end{pmatrix} =
\begin{pmatrix} 1/\mu + p_k & e^{x_k} \\ -e^{-x_k} & 0 \end{pmatrix}
\begin{pmatrix} \psi_k \\ \phi_k \end{pmatrix},
\end{equation}
\begin{equation}\label{eq: BT zcr 2}
\begin{pmatrix}\widetilde{\psi}_k \\ \widetilde{\phi}_k \end{pmatrix}=
\begin{pmatrix} 1-\lambda/\mu -\lambda^2 e^{x_k-\wx_{k-1}} & -\lambda e^{x_k} \\
\lambda e^{-\wx_{k-1}} & 1 \end{pmatrix}
\begin{pmatrix} \psi_k \\ \phi_k \end{pmatrix}.
\end{equation}
Denoting the $2\times 2$ matrices appearing in (\ref{eq: BT zcr 1}), (\ref{eq: BT zcr 2}), by $L_k(\mu)$, $V_k(\mu)$, respectively, one arrives at the zero curvature  representation of the map $F_\lambda$ with the spectral parameter $\mu$:
\[
\widetilde{L}_k(\mu)V_k(\mu)=V_{k+1}(\mu)L_k(\mu),
\]
which, in particular, yields that the trace of the monodromy matrix
\[
T(\mu)=L_N(\mu)\ldots L_2(\mu)L_1(\mu)
\]
is an integral of motion (or, better, a generating function of integrals of motion) of the map $F_\lambda$. By the way, in the open-end case one easily finds  with the help of continued fractions for $e^{\wx_k-x_k}$ given in (\ref{eq: BT Toda expl}) that
\[
\prod_{k=1}^N e^{\wx_k-x_k}=\lambda^N(T(\lambda))_{11}.
\]
In the periodic case, one has to replace the right-hand side by one of the eigenvalues of the matrix $\lambda^N T(\lambda)$. In both cases, the zero-curvature representation yields that the left-hand side is an integral of motion. Recall that this statement is equivalent to the discrete Lagrangian 1-form being closed on solutions.

%%%%%%%%%%%%%%%%%%%%%%%%%%%%%%%
%%%%%%%%%%%%%%%%%%%%%%%%%%%%%%%
\section{Conclusions}
In this paper, we developed a variational characterization of systems of commuting Hamiltonian flows (resp. of commuting symplectic maps). This characterization can be formulated as extremizing the action obtained by integration of a Lagrangian 1-form along an {\em arbitrary} curve in the multi-dimensional time. There remains a challenge to develop a similarly complete theory for two- and three-dimensional systems, both in the continuous and discrete contexts. This is the subject of the ongoing research.

This research is supported by the DFG Collaborative Research Center TRR 109 ``Discretization in Geometry and Dynamics''.
%%%%%%%%%%%%%%%%%%%%%%%%%%%%%%%
%%%%%%%%%%%%%%%%%%%%%%%%%%%%%%%

%%%%%%%%%%%%%%%%%%%%%%%%%%%%%%%
%%%%%%%%%%%%%%%%%%%%%%%%%%%%%%%

%%%%%%%%%%%%%%%%%%%%%%%%%%%%%%%
%%%%%%%%%%%%%%%%%%%%%%%%%%%%%%%
\end{document}